%% file: XabierCid_LHCP.tex
\documentclass[10pt]{article}
\usepackage{graphicx}

\def\Title#1{\begin{center} {\Large #1 } \end{center}}
\def\Author#1{\begin{center}{ \sc #1} \end{center}}
\def\Address#1{\begin{center}{ \it #1} \end{center}}

\newcommand\pubblock{\rightline{\begin{tabular}{l} Proceedings of the Fifth Annual LHCP\\ 
         \pubdate  \end{tabular}}}

\newenvironment{Abstract}{\begin{quotation} \begin{center} 
             \large ABSTRACT \end{center}\bigskip 
      \begin{center}\begin{large}}{\end{large}\end{center} \end{quotation}}

\newenvironment{Presented}{\begin{quotation} \begin{center} 
             PRESENTED AT\end{center}\bigskip 
      \begin{center}\begin{large}}{\end{large}\end{center} \end{quotation}}

\input econfmacros.tex

\usepackage{color}

\newcommand\pubdate{\today}

\def\affiliation{
On behalf of the LHCb Collaboration, \\
Departamento de F\'{i}sica de Part\'{i}culas (USC) \\ R\'{u}a de Xoaqu\'{i}n Díaz de R\'{a}bago, s/n.\\
Santiago de Compostela, Spain}

\usepackage{ifthen,placeins,lineno,hyperref}

\newboolean{articletitles}
\setboolean{articletitles}{true} 

\newboolean{inbibliography}
\setboolean{inbibliography}{true}

\newboolean{uprightparticles}
\setboolean{uprightparticles}{false} 

\input{preamble}

\begin{document}

\large
\begin{titlepage}
\pubblock

\vfill
\Title{  Physics with jets in LHCb }
\vfill

\Author{ Xabier Cid Vidal  }
\Address{\affiliation}
\vfill
\begin{Abstract}

LHCb, while purpose built for $b-$physics, also functions as a general purpose forward detector, covering the pseudo-rapidity range 2.0 to 5.0. LHCb has performed several measurements including jets, which concern, \eg, QCD, top and Higgs physics. A selection of LHCb results in this area will be presented,
focusing on the most recent ones. 

\end{Abstract}
\vfill

\begin{Presented}
The Fifth Annual Conference\\
 on Large Hadron Collider Physics \\
Shanghai Jiao Tong University, Shanghai, China\\ 
May 15-20, 2017
\end{Presented}
\vfill
\end{titlepage}
\def\thefootnote{\fnsymbol{footnote}}
\setcounter{footnote}{0}
%

\normalsize 


\section{Introduction}

LHCb is one of the four largest detectors of the Large Hadron Collider (LHC) project at CERN (Geneva, Switzerland). It is a single-arm forward spectrometer with a fully instrumented detector with a unique coverage in terms of pseudo-rapidity:  $2<\eta<5$ \cite{lhcb_detector}. While originally designed to study the production and decay of $b$ and $c$ hadrons, LHCb has extended its physics programme to also include other areas such as physics with jets. 

The results presented in these proceedings correspond to proton-proton collision data taken by LHCb during Runs 1 and Run 2 of the LHC, recorded at centre of mass energies of $\sqrt{s}$ = 8 TeV (during the year 2012) and 13 TeV (during the year 2016).
LHCb recorded 2 \invfb of data in 2012 and 1.7 \invfb of data in 2016.

Measurements of jets at LHCb address several interesting areas, such as:
\begin{itemize}
\item QCD: to set important constraints on proton PDFs, to  probe hard QCD in a unique kinematic range or to look into jet properties.
\item Higgs physics: in direct searches for the Higgs boson decaying to $b\bar{b}$ and $c\bar{c}$ final states.
\item Direct searches of long-lived beyond the SM particles decaying to jets.
\end{itemize} 

In these proceedings, the first two groups of topics will be covered, while examples of the third can be found elsewhere \cite{martino}.

\section{Jet reconstruction at LHCb}

Jets are reconstructed at LHCb using using a particle flow algorithm \cite{LHCb-PAPER-2013-058} and clustered using the \antikt algorithm \cite{antikt} with $R = 0.5$. The calibration of the jet reconstruction is performed in data using $\PZ \to \mup\mun$ events containing a jet reconstructed back-to-back with respect to the \PZ. The efficiency for reconstructing and identifying jets is around 90\% for jets with a transversal momentum, \pt, above 20 \gevc. 

\begin{figure}[htb]
\begin{center}
\includegraphics[width=0.28\textwidth]{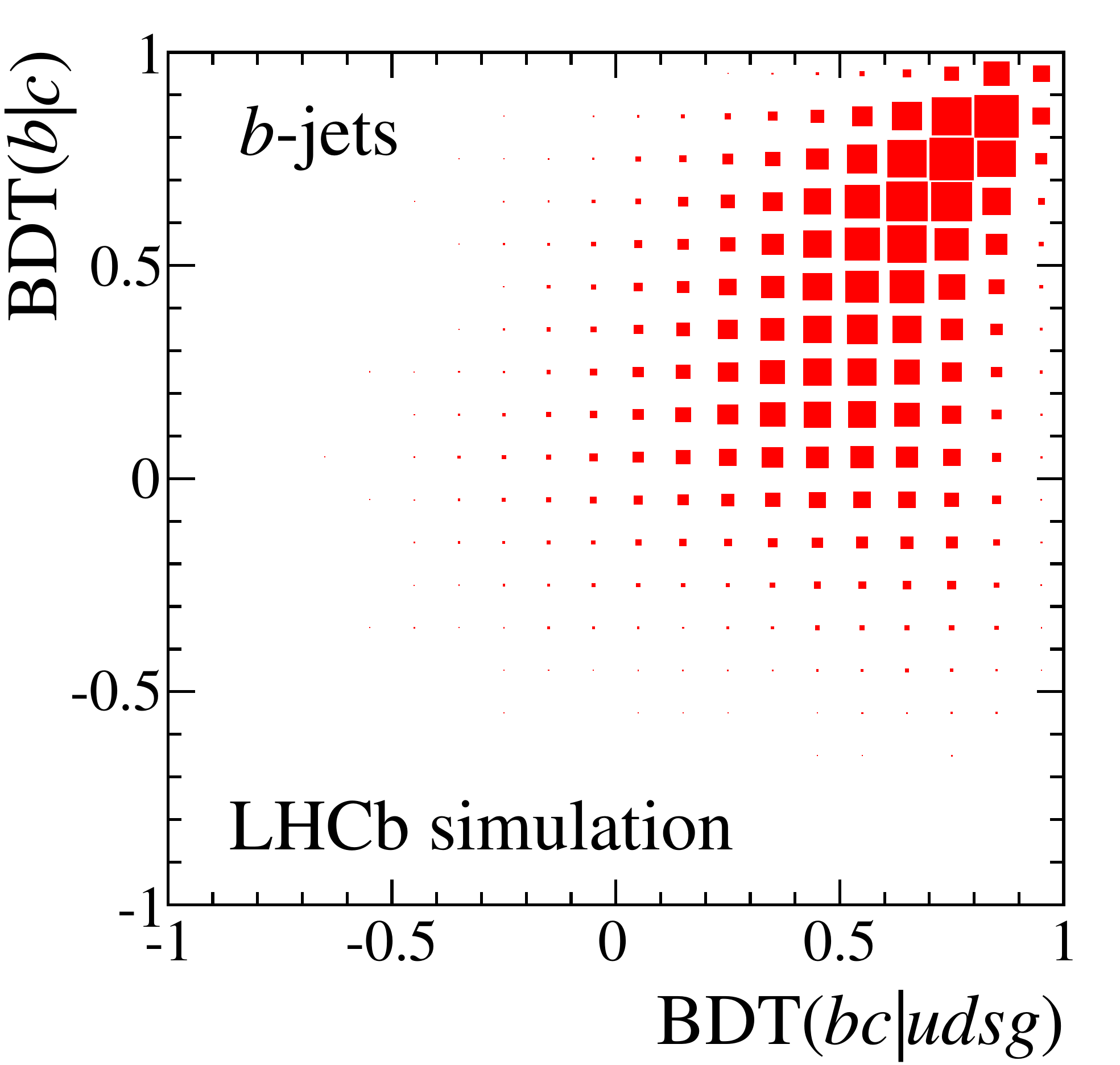}
\includegraphics[width=0.28\textwidth]{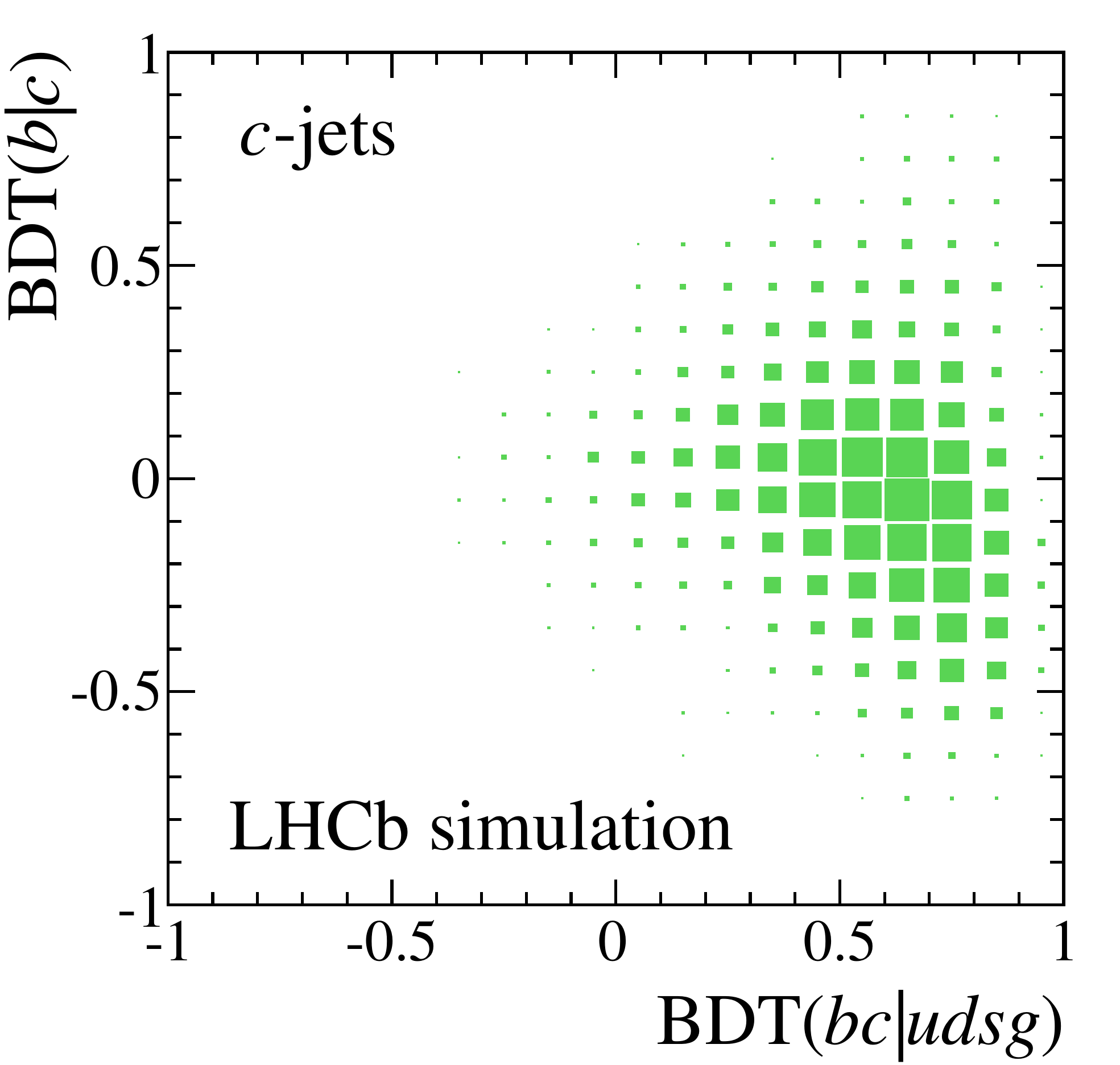}
\includegraphics[width=0.4\textwidth]{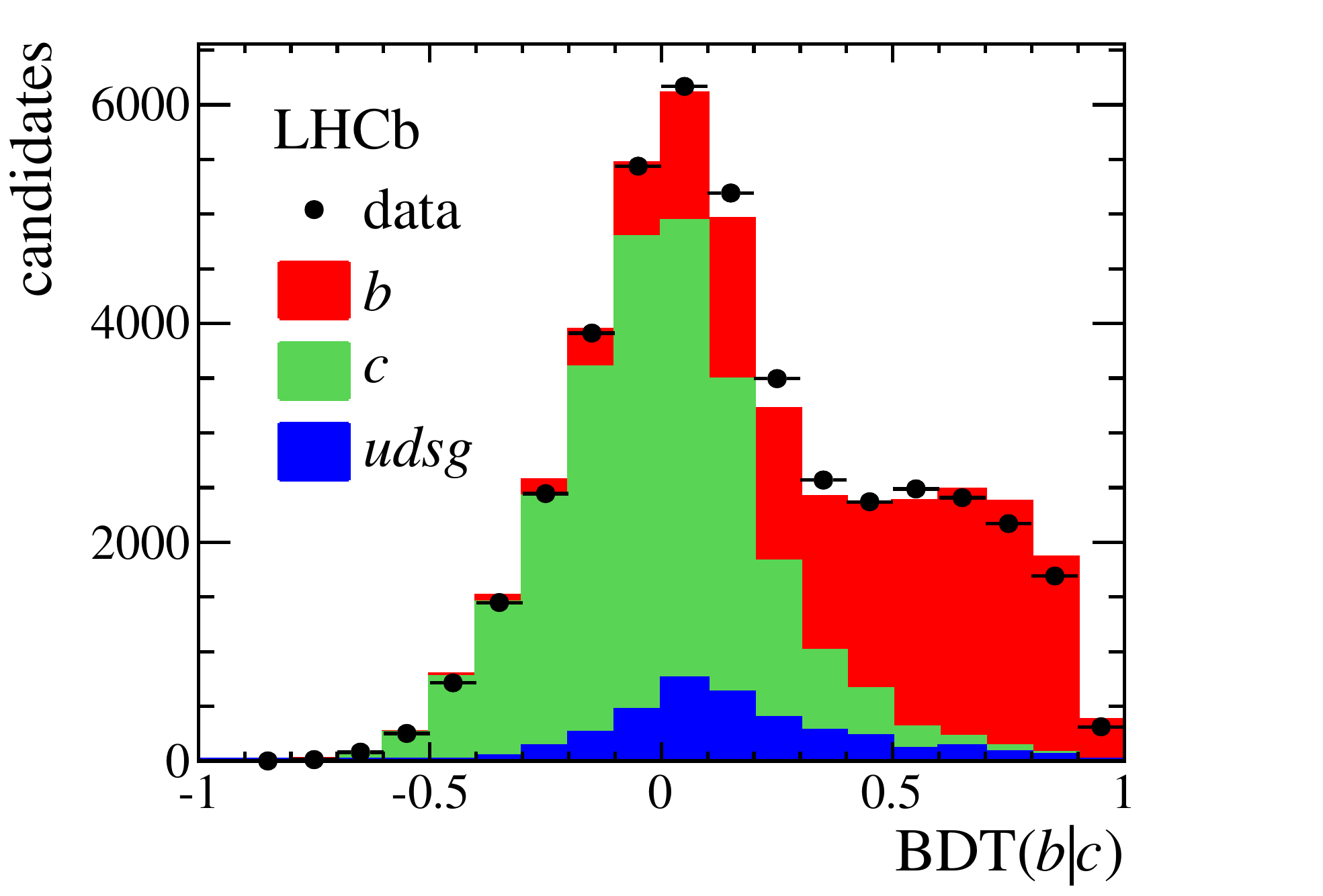}
\caption{Two-dimensional distribution of \bdtbc vs.~\bdth for MC $b$ (left) and $c$ (center) jets. \bdtbc distribution in a $c-$enriched data sample. More details about the method can be found in \cite{LHCb-PAPER-2015-016}.  \label{fig:tag} }
\end{center}
\end{figure}

Furthermore, LHCb has developed  \cite{LHCb-PAPER-2015-016} a method to tag jets and determine whether they correspond to a $b$ or $c$ hadron or to a lighter particle. Jets are tagged whenever a secondary vertex (SV) is reconstructed close enough to the jet in terms of $\Delta R = \sqrt{\Delta \phi^2+\Delta \eta^2}$. This provides a  light jet mistag rate below 1\%, with an efficiency on $b$ ($c$) jets of $\sim65$\% ($\sim25$ \%). Moreover, using the SV and jet properties, two BDTs \cite{tmva} have been developed: one to separate heavy from light jets (\bdth) and one to separate $b$ from $c$ jets (\bdtbc). Figure \ref{fig:tag} shows the two-dimensional distribution of \bdtbc vs.~\bdth for MC $b$ and $c$ SV-tagged jets, together with the \bdtbc distribution in a $c-$enriched data sample. Fitting the \bdtbc distribution allows separating the different components in data.

\section{Measurements with EW bosons and jets}

\subsection{\texorpdfstring{\PZ and \PW + jets production}{Z and W + jets production}}\label{sec:wjets}

LHCb has measured the production of \PZ and \PW bosons associated with jets using the data collected at $\sqrt{s}=8$ TeV \cite{wjets}. Jets are reconstructed as explained above, while \PZ and \PW bosons are reconstructed using the muonic final states. The production of \PW boson plus jets is discriminated from misidentified background processes arising in QCD using a muon isolation variable, which is built as the ratio between the \pt of the jet containing the muon and the \pt of the muon alone. Figure \ref{sec:wjets} (left) shows the distribution of this variable, with genuine muons from the \PW boson peaking at 1. 

\begin{figure}[htb]
\begin{center}
\includegraphics[width=0.53\textwidth]{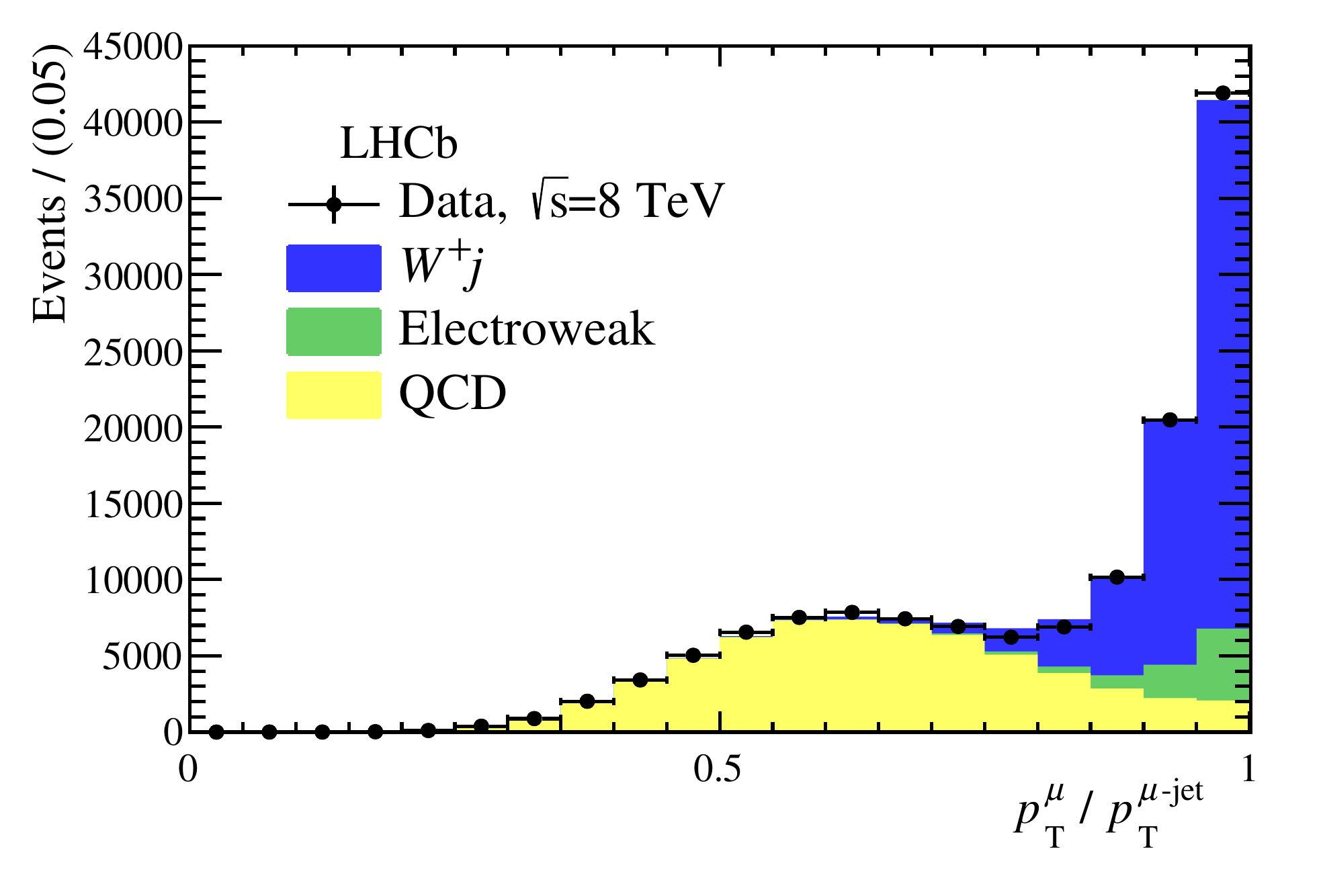}
\includegraphics[width=0.35\textwidth]{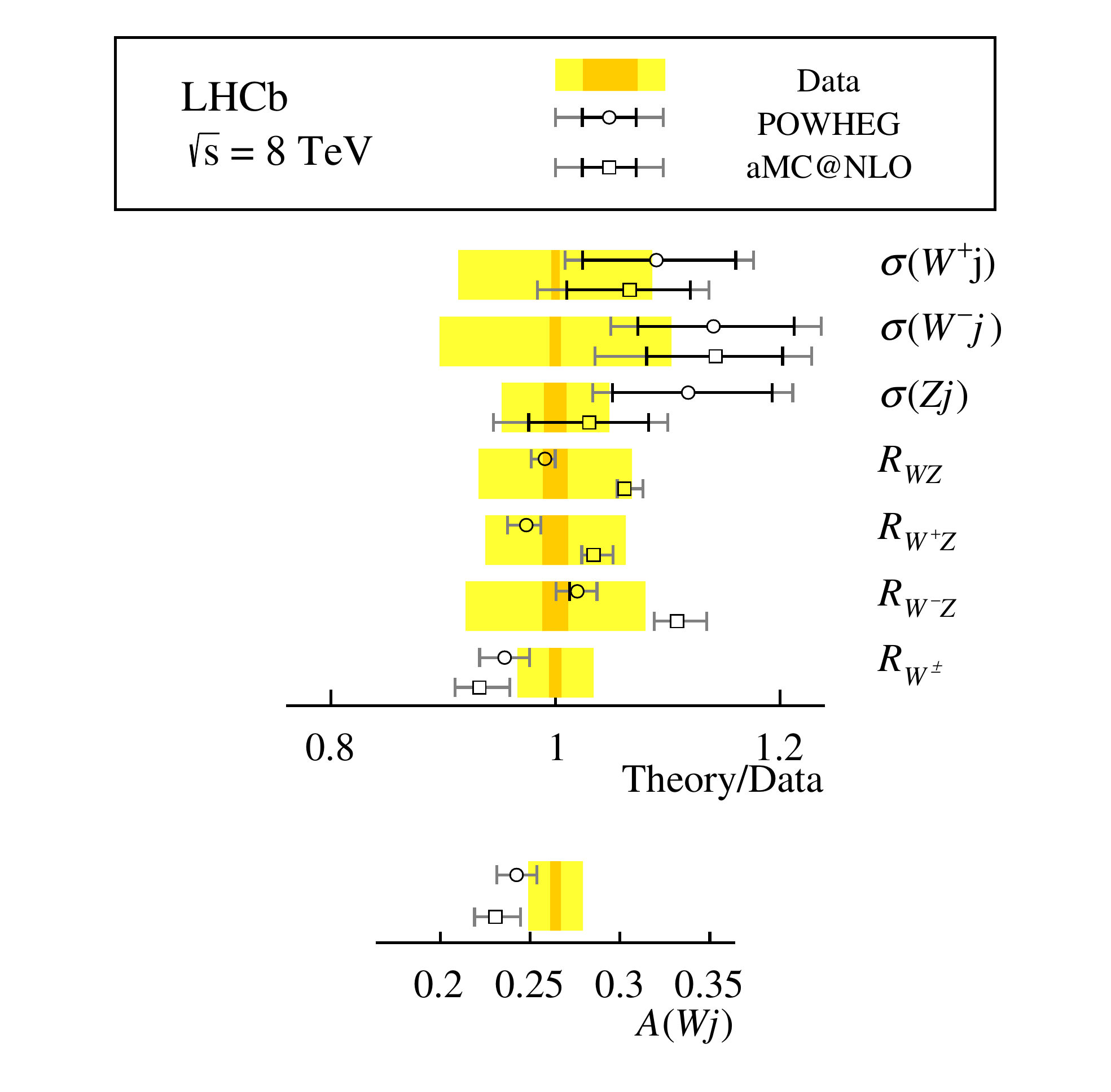} \hspace{0.5cm}
\caption{Left, muon isolation, which allows to separate muons from \PW from QCD background.
Right, measurement of absolute cross sections of \PZ and \PW bosons plus jets, ratios between those ($R$) and charge asymmetry ($A$) between $W^+$ and $W^-$ plus jets (the cross-sections and ratios are shown normalised to the measurement, while the asymmetry is presented separately) \cite{wjets}.  \label{fig:w} }
\end{center}
\end{figure}

The absolute and differential cross sections, as well as ratios of these and charge asymmetry have been measured in this analysis and compared to different theoretical predictions. The agreement found is good in general. A graphical example can be found in Fig.~\ref{fig:w} (right picture). 

\subsection{\texorpdfstring{\wpb, \wpc and \ttbar}{W+bb, cc and ttbar}} \label{sec:wbb}

LHCb has measured the \wppb, \wppc, \wmpb, \wmpc and \ttbar production cross sections using a sample of pp collisions taken at $\sqrt{s}=8$ TeV with a high-\pt isolated lepton (electron or muon) and two heavy flavour ($b$ or $c$) tagged jets in the final state. The channel \wpc is studied for the first time. 

\begin{figure}[htb]
\centering
\begin{minipage}[c]{0.08\textwidth}
\includegraphics[width=\textwidth]{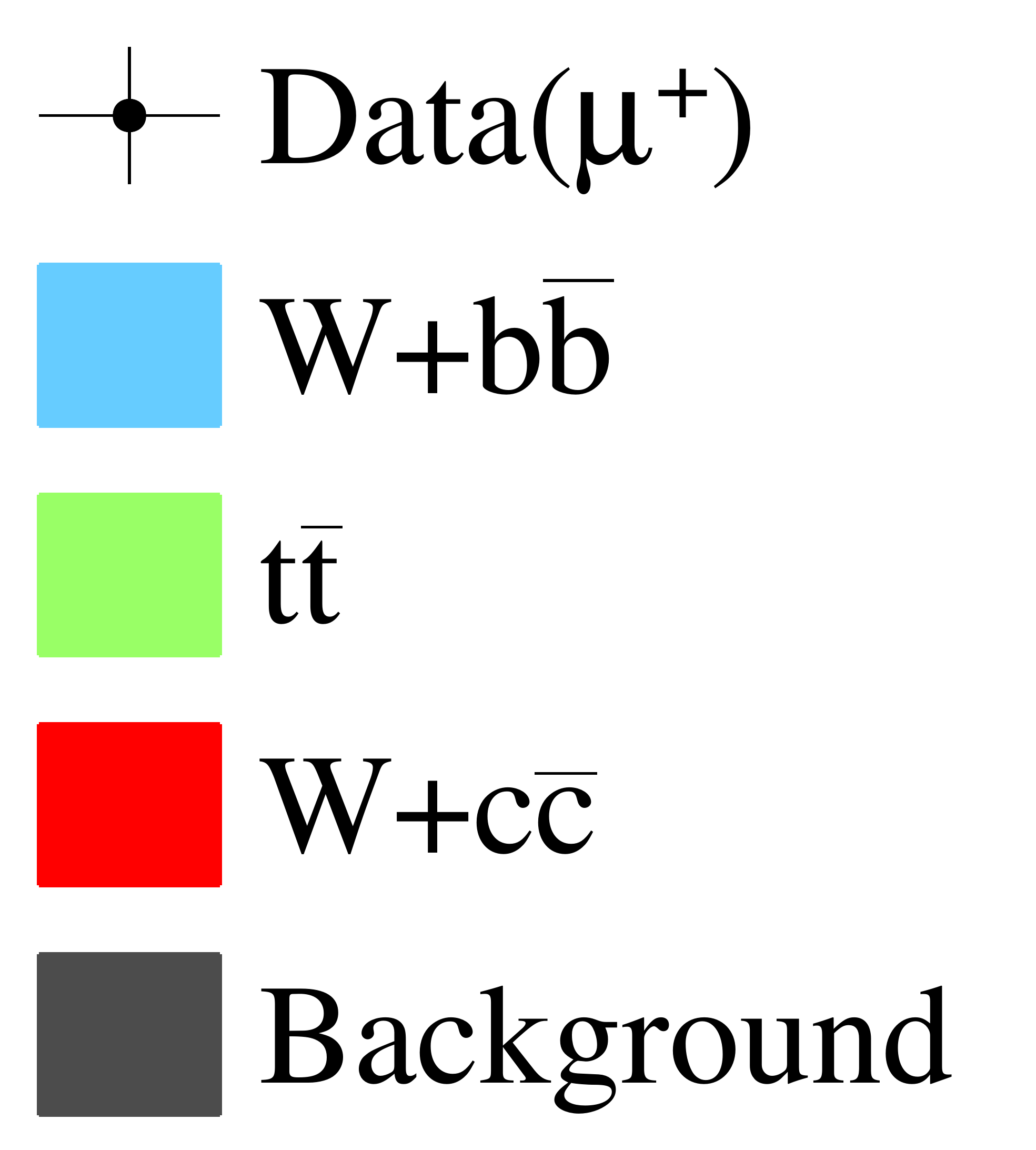}
\end{minipage}
\begin{minipage}[c]{0.37\textwidth}
\includegraphics[width=\textwidth]{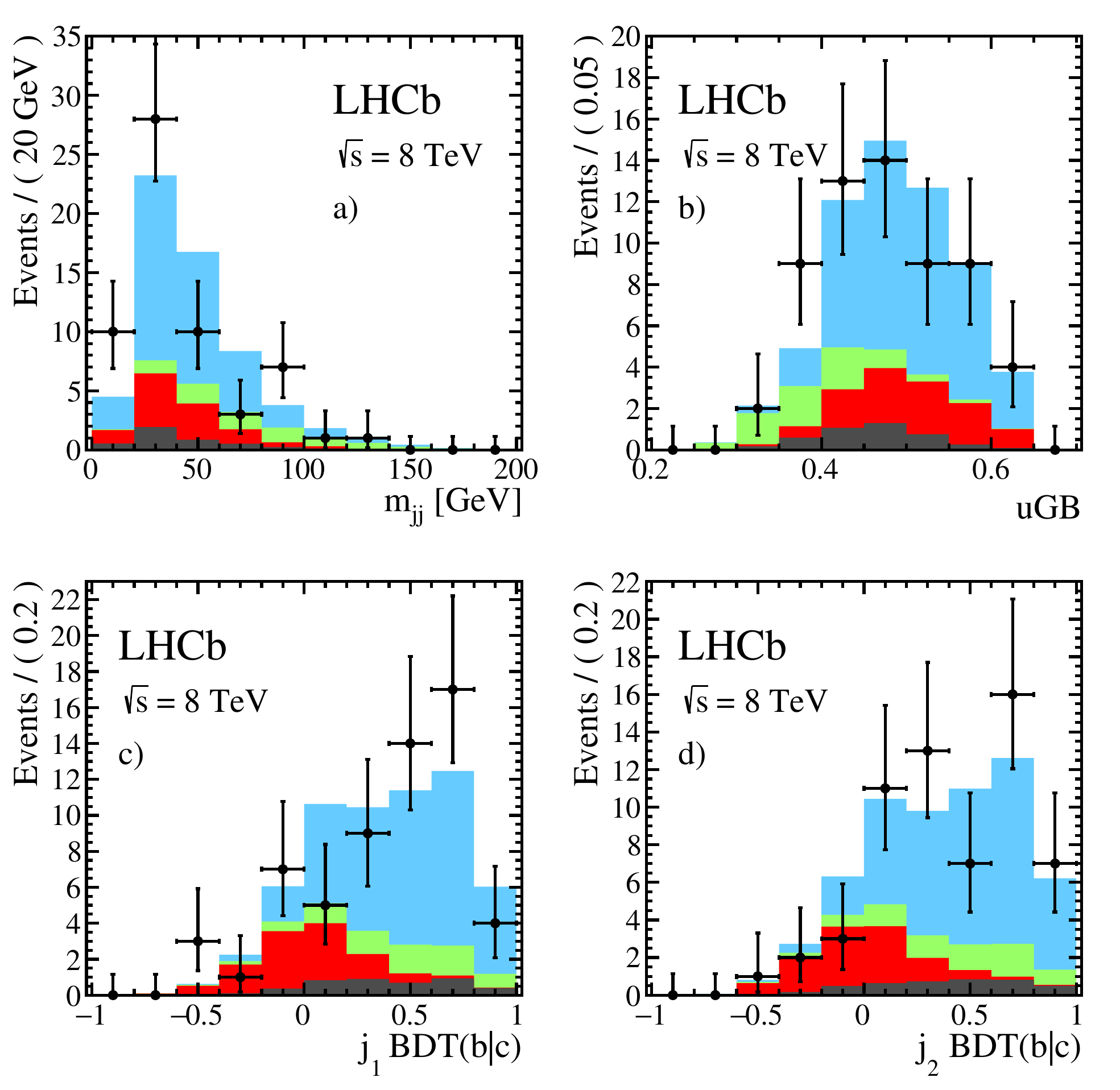}
\end{minipage}
\begin{minipage}[c]{0.53\textwidth}
\includegraphics[width=\textwidth]{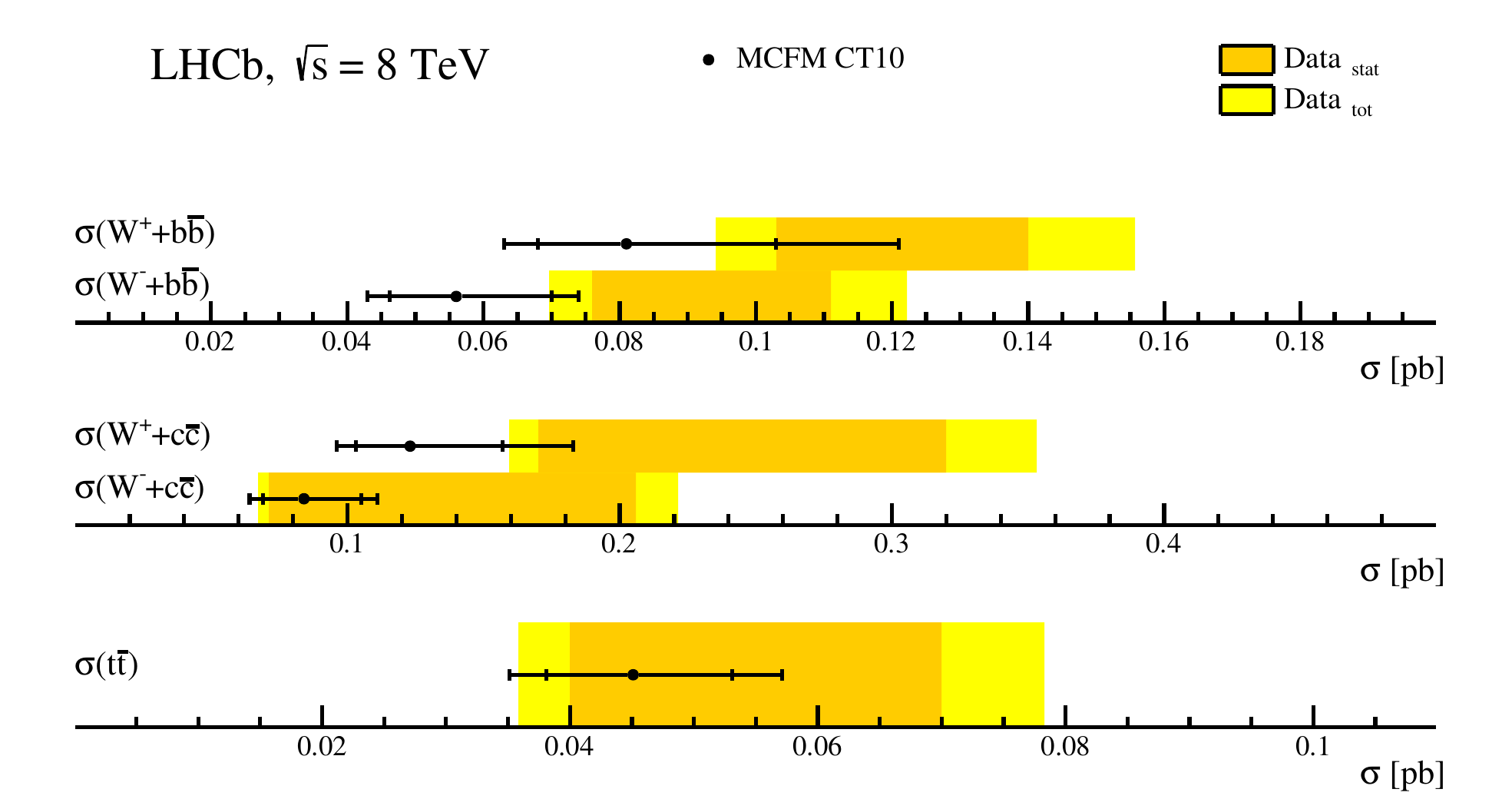}
\end{minipage}
\caption{Left: Example of the fit used to extract the production cross sections in \cite{wbb}. Projections of the simultaneous 4D-fit results are shown for the $\mu^+$ sample corresponding to: a) the dijet mass; b) the \ugb response; the \bdtbc of the c) leading and d) sub-leading jets. Right: Graphical display of the measured \wpb, \wpc and \ttbar production cross sections and the corresponding NLO theoretical SM predictions. 
\label{fig:wbb}}
\end{figure}

In this analysis, jets are reconstructed as already described and those coming from heavy quarks tagged using the method explained above. In order to extract the different signal components, a simultaneous 4-D fit is performed to the \mup, \mun, \ep and \en samples.
The four variables used in the fit are the dijet mass, a MVA-response to discriminate \ttbar from \wpb and \wpc events (referred to as \ugb \cite{ugb})  and \bdtbc, used for both jets. As an example, Fig.~\ref{fig:wbb} (left picture) shows the projections of this fit for the \mup sample.

This analysis measures the production of \wppb, \wppc, \wmpb, \wmpc and \ttbar with statistical significances of 7.1$\sigma$, 4.7$\sigma$, 5.6$\sigma$, 2.5$\sigma$ and 4.9$\sigma$ respectively. The cross sections obtained in the LHCb fiducial region and the Next-to-Leading-Order (NLO) theory predictions can be found in Fig.~\ref{fig:wbb} (right picture).

\subsection{\texorpdfstring{Search for associated production of the Higgs boson decaying to $b\bar{b}$}{Search for associated production of the Higgs boson decaying to bb}}

\begin{figure}[htb]
\centering
\includegraphics[width=0.44\textwidth]{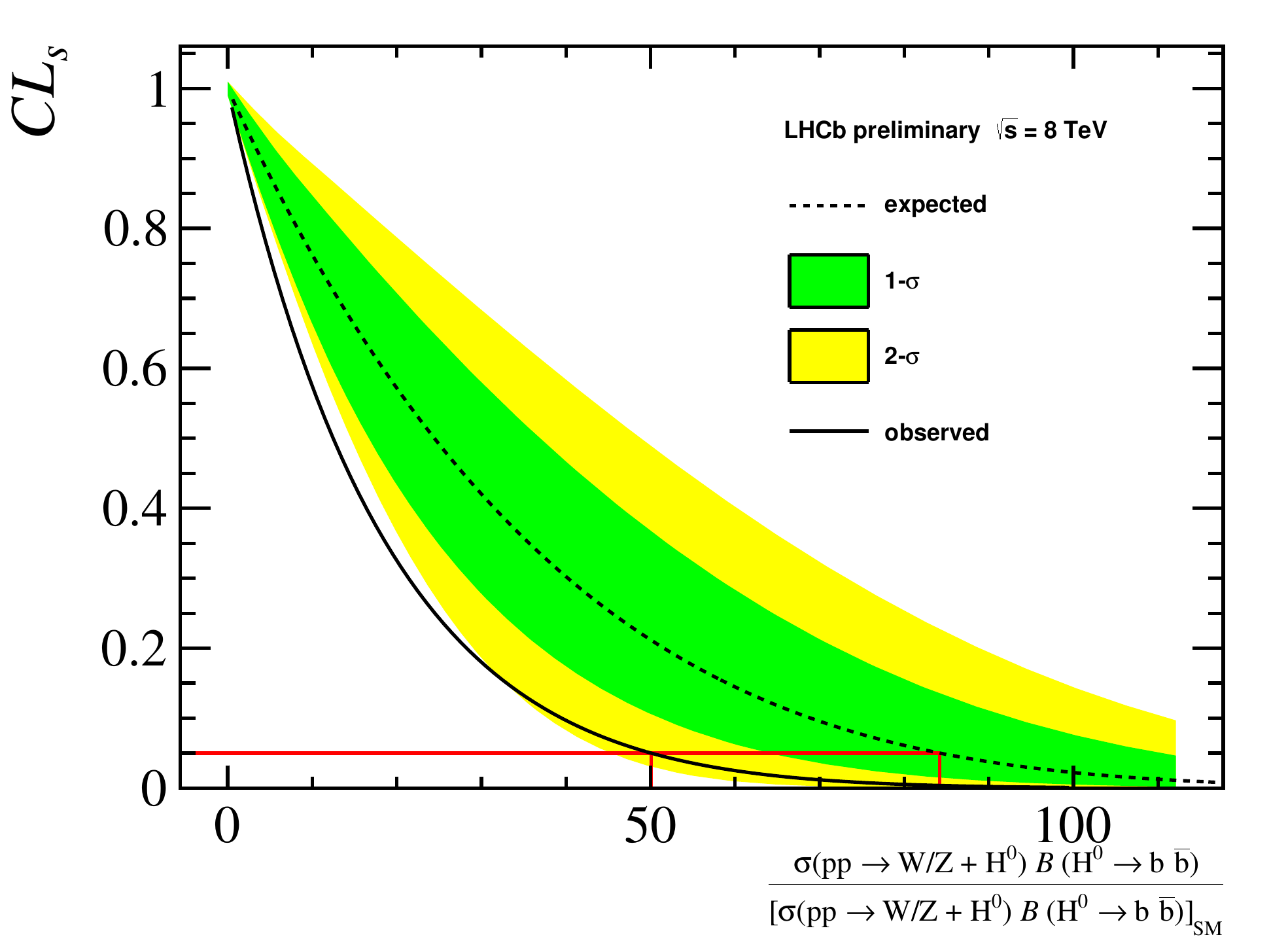}
\includegraphics[width=0.47\textwidth]{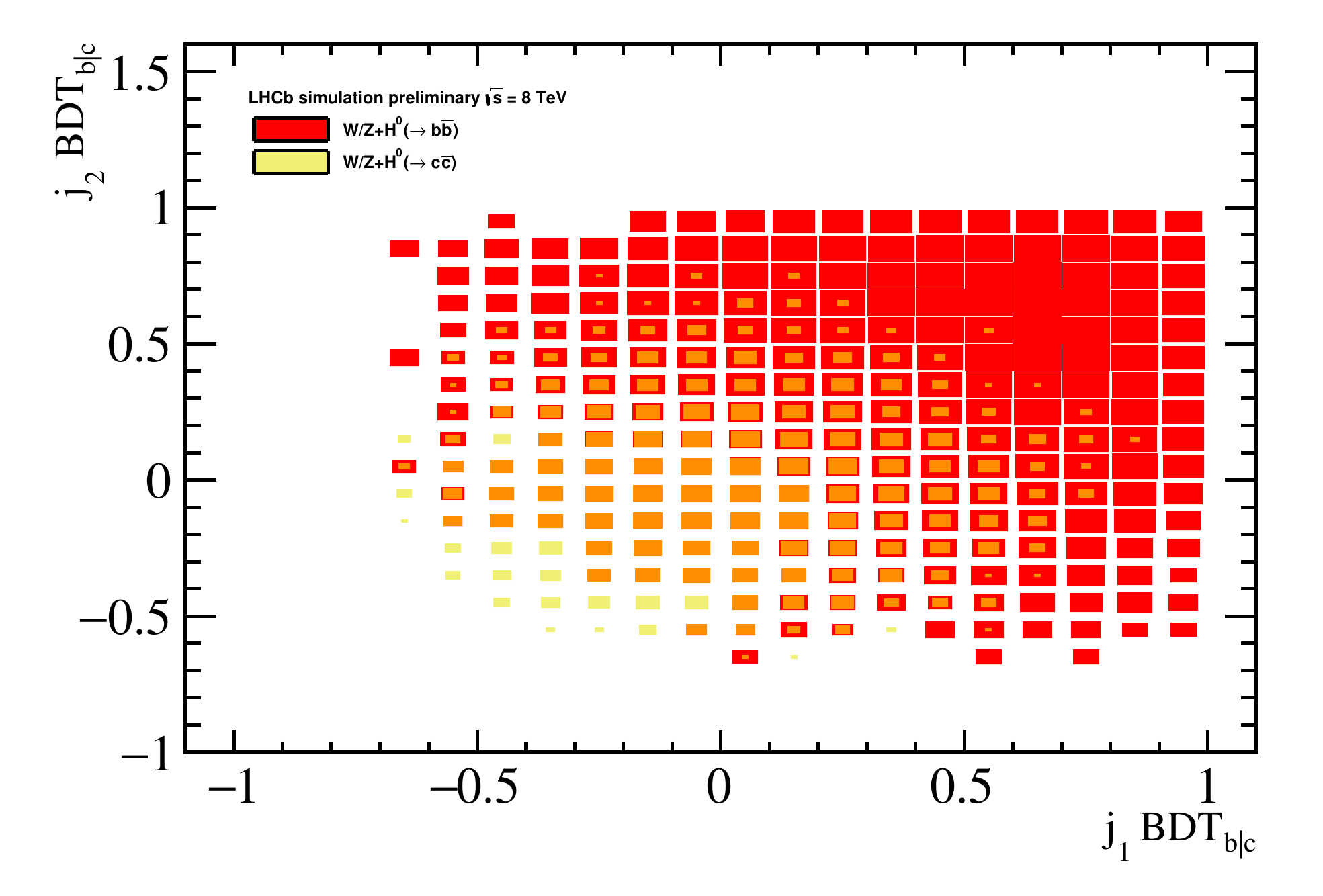}
\caption{Left, $\cls$ vs.~the product $\BR(\hbb) \times \sigma(pp \to \PW/\PZ + \H)$ normalized to the SM values. This plot was used to obtain a 95\% CL upper limit on this ratio of $\sim50$. Right, \bdtbc distribution for both jets in \hbb (red) and \hcc (yellow) decays. The relative normalization corresponds to the SM branching fractions. More information can be found in \cite{LHCb-CONF-2016-006}.  \label{fig:higgs} }
\end{figure}

Using the same dataset as in Sect.~\ref{sec:wbb}, LHCb has performed
search for the SM Higgs boson decaying to a pair of \bbbar quarks and produced in association with a \PW or \PZ boson
 \cite{LHCb-CONF-2016-006}. In this case, the cut on the \pt of the jets is tightened to reduce the \wpb background, and two \ugb are
 created to discriminate signal from background: one to separate signal from \ttbar and another to separate signal from \wpb. 
 Then, the expected background is fixed to the theory predictions and this is compared to data using the \cls method. The result obtained,
 showed in Fig.~\ref{fig:higgs} is not competitive with ATLAS and CMS, but it is currently the best limit on Higgs boson production at LHCb, and shows the future potential in this channel.
 
 At the same time, as part of this analysis, a search for \hcc was performed too. Although the final result was not significant, a search
 could be performed if a High Luminosity LHCb upgrade takes place. As an example of the LHCb discrimination capabilities 
 between \hbb and \hcc, Fig.~\ref{fig:higgs} shows the \bdtbc distribution for both jets in both decays, normalized to their relative SM
 branching fractions.

\section{\texorpdfstring{\jpsi production in jets}{Jpsi production in jets}}

Measuring the production of \jpsi within jets can be useful to study QCD phenomenology. In this regard,
for instance, \jpsi are expected to be isolated if they are produced directly in parton-parton scattering.
To study this, \zjpsi is defined as the ratio of the \pt of the \jpsi and the \pt of the jet that contains the \jpsi
($\zjpsi\equiv\pt(\jpsi)/\pt(jet)$). Therefore, \zjpsi is a good proxy of the \jpsi isolation.

\begin{figure}[htb]
\begin{center}
\includegraphics[width=0.45\textwidth]{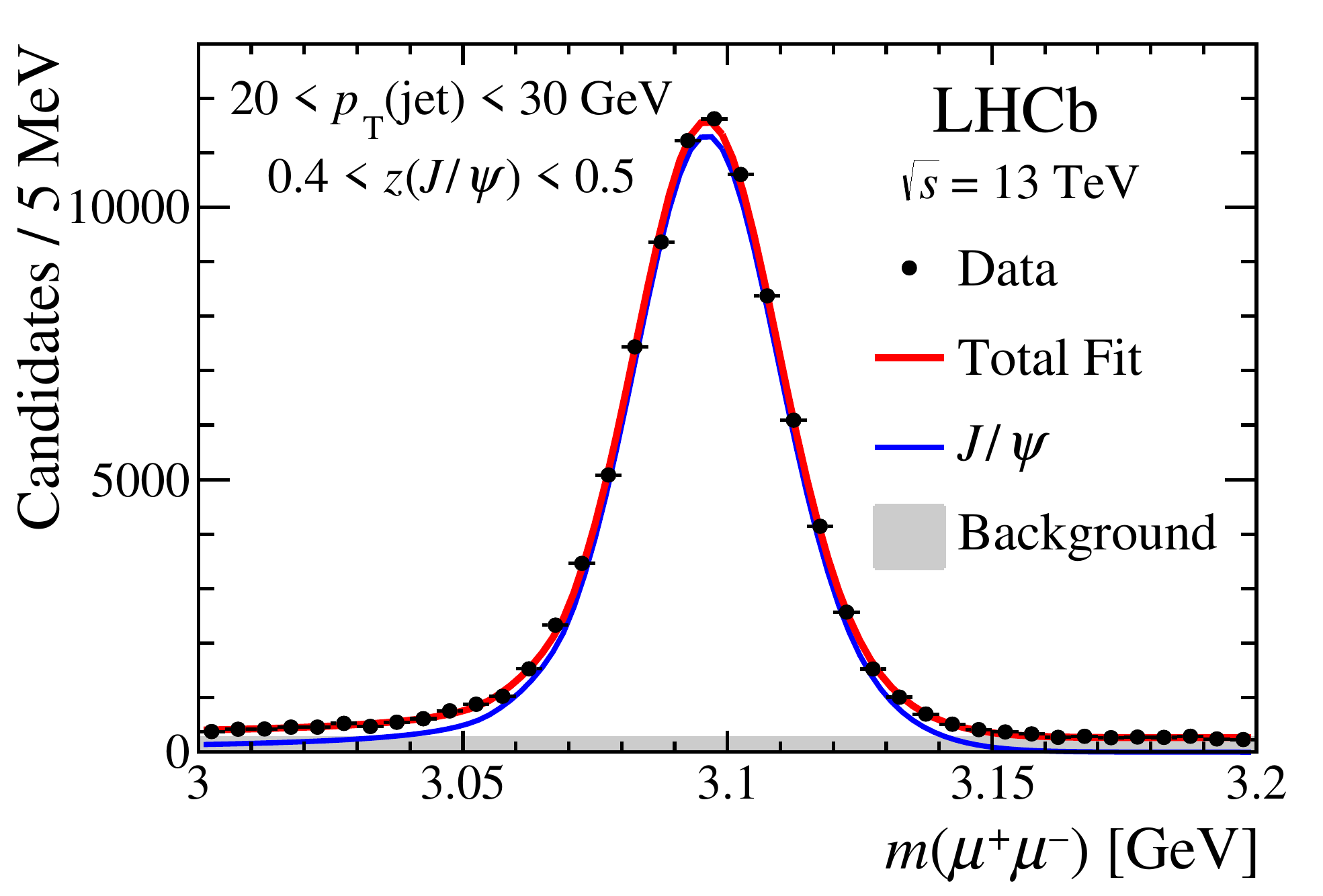}
\includegraphics[width=0.45\textwidth]{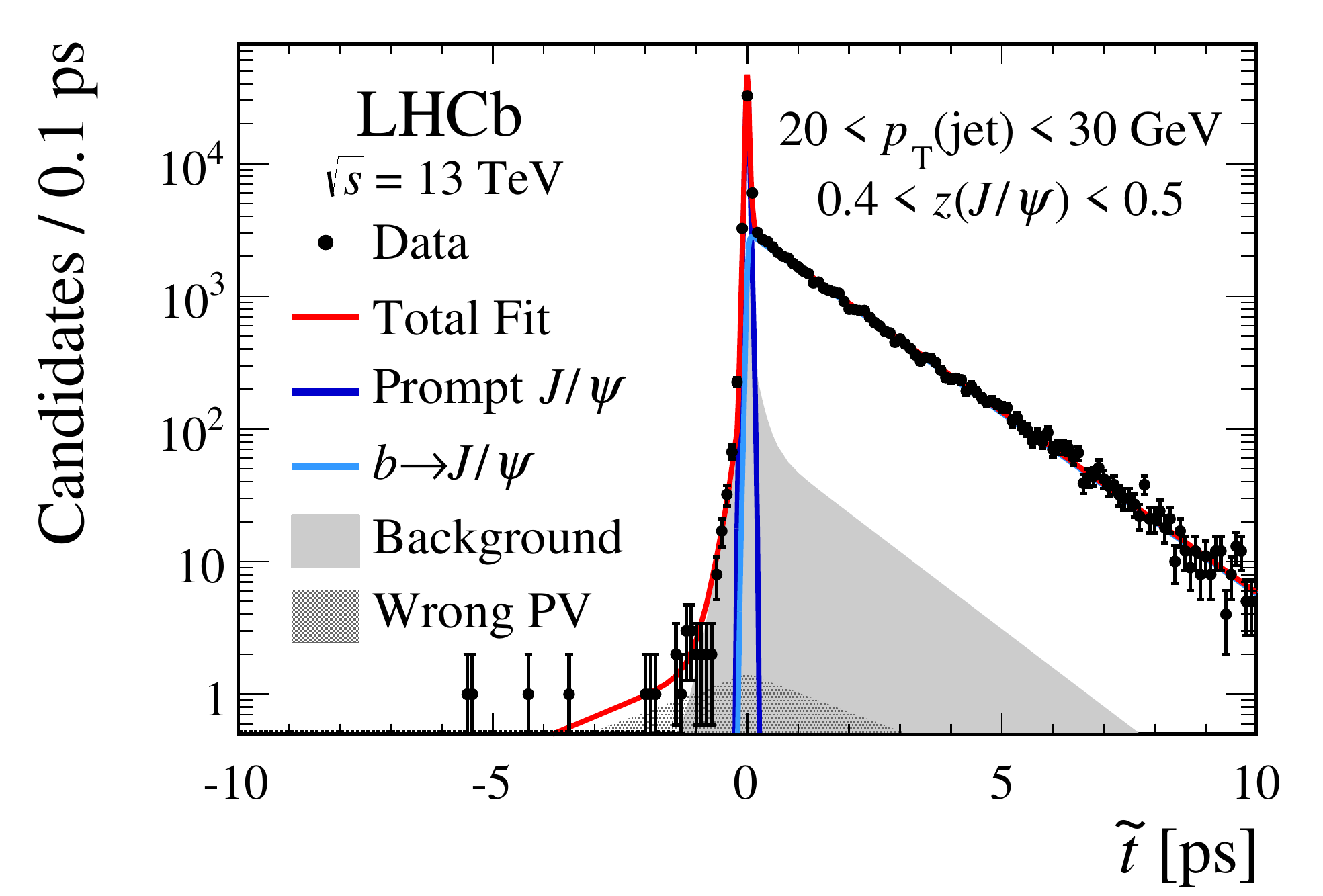}
\caption{Examples of fits to subtract the \jpsi yields \cite{jpsi}. Left, dimuon invariant mass. Right, pseudo-lifetime, which allows to differenciate prompt \jpsi from those coming from a $b$ quark decay. \label{fig:jpsi1} }
\end{center}
\end{figure}

LHCb has measured the distribution of \zjpsi using 2016 data taken at $\sqrt{s}=13$ TeV \cite{jpsi}. It should be highlighted
that for this analysis the \jpsi were selected directly at trigger level, which allowed a significant increase of statistics.
After this, the jets are reconstructed offline, and the \jpsi are separated if they are prompt or produced from a $b$ quark decay (detached). For this,
a pseudo-lifetime is defined and a fit is performed to separate both components. Figure \ref{fig:jpsi1} shows this fit, together with an example
of the dimuon invariant mass fits used to discriminate \jpsi from the combinatorial background.

In order to measure the raw distribution of \zjpsi, the detector response has to be corrected. For this, 
an iterative 2D Bayesian unfolding is performed, which corrects 
for the \zjpsi  and $\pt(jet)$ resolution, which are at the level of $\sim20 - 25$\%.

\begin{figure}[htb]
\begin{center}
\includegraphics[width=0.45\textwidth]{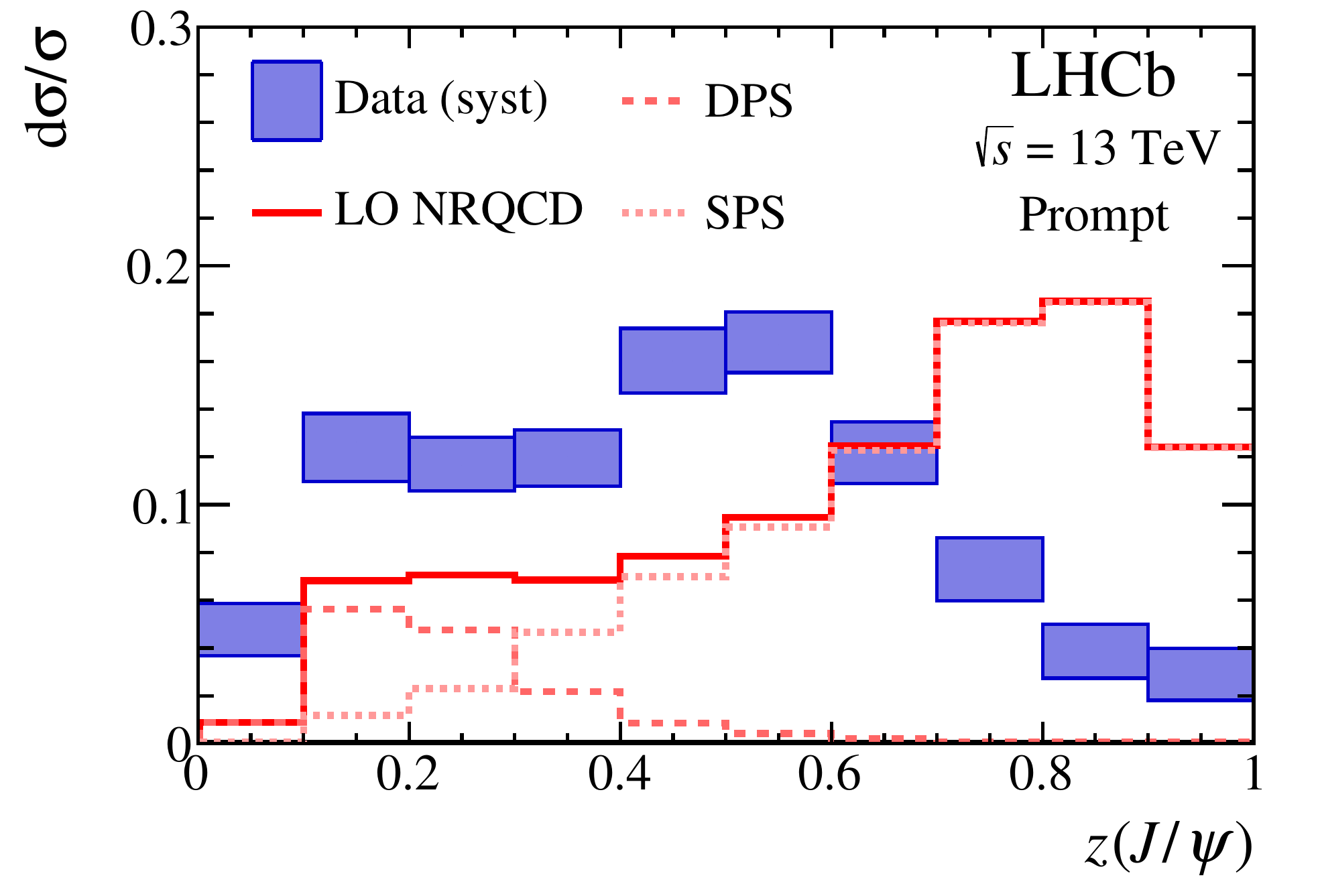}
\includegraphics[width=0.45\textwidth]{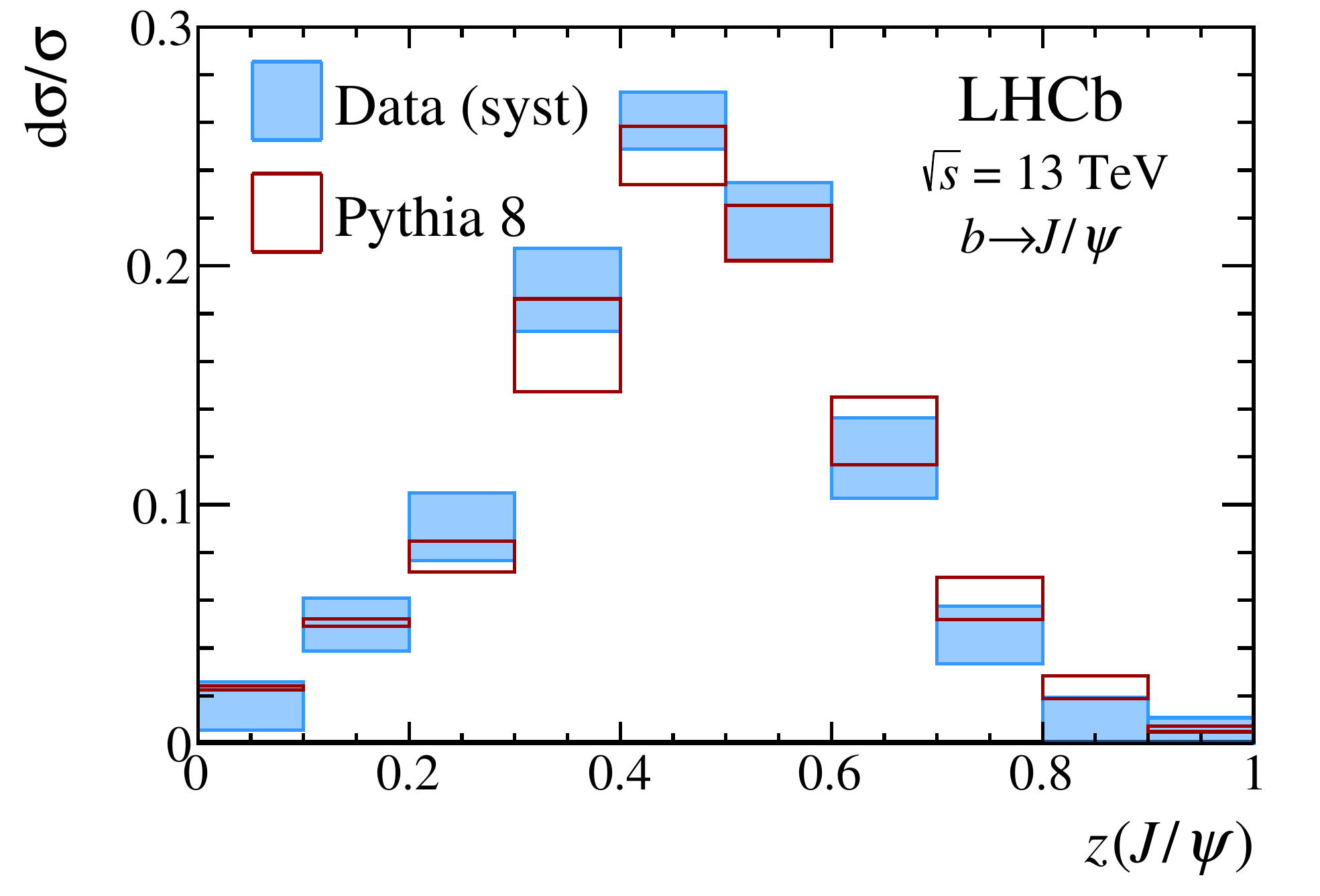}
\caption{$\zjpsi\equiv\pt(\jpsi)/\pt(jet)$ distribution for prompt \jpsi (left) and \jpsi coming from a $b$ quark decay (right) \cite{jpsi}. In both cases, the results are compared to the theory predictions. \label{fig:jpsi2} }
\end{center}
\end{figure}

With all this, the \zjpsi distribution is measured for prompt and detached \jpsi.
The results are compared to the \pythia8 predictions. The results can be found in Fig.~\ref{fig:jpsi2}. Although
the detached \zjpsi distribution is well described by \pythia8 \cite{pythia}, this is not the case with the prompt one, even when Double Parton Scattering effects are taken into account. This result indicates that additional processes must be considered to explain \jpsi production, either through directly considering higher order effects in the hard process, or through simulating \jpsi production in the parton shower. In this regard,
 as a consequence of this result, alternative descriptions of the quarkonium production have already been suggested,
providing a better qualitative description of the prompt \zjpsi distribution \cite{jpsi_th}.

\section{Conclusions}

Although it was initially designed for flavour physics, LHCb can now be considered a general purpose detector in the forward direction. Several examples of this have been shown through results involving jets. The Run 2 of the LHC will extend even more the capabilities of LHCb beyond flavour physics.

\FloatBarrier

\input{XabierCid_LHCP.bbl}
\end{document}

%% file: econfmacros.tex
\def\beq{\begin{equation}}
\def\eeq#1{\label{#1}\end{equation}}
\def\eeqn{\end{equation}}

\def\beqa{\begin{eqnarray}}
\def\eeqa#1{\label{#1}\end{eqnarray}}
\def\eeqan{\end{eqnarray}}

\let\bar=\overbar

\def\eg{{\it e.g.}}

\def\W{{\cal W}}

\def\Dslash{\not{\hbox{\kern-4pt $D$}}}
\def\dslash{\not{\hbox{\kern-2pt $\del$}}}

\def\BR{\mbox{\rm BR}}

\def\msb{{\bar{\ssstyle M \kern -1pt S}}}



%% file: preamble.tex
\usepackage{microtype}
\usepackage{lineno}  
\usepackage{xspace} 
\usepackage{caption} 

\usepackage{graphicx}  
\usepackage{color}
\usepackage{colortbl}
\graphicspath{{./figs/}} 

\usepackage{amsmath} 
\usepackage{amssymb}
\usepackage{amsfonts}
\usepackage{upgreek} 

\newcommand*\patchAmsMathEnvironmentForLineno[1]{%
\expandafter\let\csname old#1\expandafter\endcsname\csname #1\endcsname
\expandafter\let\csname oldend#1\expandafter\endcsname\csname
end#1\endcsname
 \renewenvironment{#1}%
   {\linenomath\csname old#1\endcsname}%
   {\csname oldend#1\endcsname\endlinenomath}%
}
\newcommand*\patchBothAmsMathEnvironmentsForLineno[1]{%
  \patchAmsMathEnvironmentForLineno{#1}%
  \patchAmsMathEnvironmentForLineno{#1*}%
}
\AtBeginDocument{%
\patchBothAmsMathEnvironmentsForLineno{equation}%
\patchBothAmsMathEnvironmentsForLineno{align}%
\patchBothAmsMathEnvironmentsForLineno{flalign}%
\patchBothAmsMathEnvironmentsForLineno{alignat}%
\patchBothAmsMathEnvironmentsForLineno{gather}%
\patchBothAmsMathEnvironmentsForLineno{multline}%
\patchBothAmsMathEnvironmentsForLineno{eqnarray}%
}

\input{lhcb-symbols-def} 

\usepackage{cite} 
\usepackage{mciteplus}

%% file: lhcb-symbols-def.tex
\def\bdtbc     {\ensuremath{{\rm BDT}(b|c)}\xspace}
\def\bdth     {\ensuremath{{\rm BDT}(bc|udsg)}\xspace}
\def\cls     {\ensuremath{{\rm CL_{S}}}\xspace}
\def\zjpsi     {\ensuremath{z(\jpsi)}\xspace}

\def\MagUp {\mbox{\em Mag\kern -0.05em Up}\xspace}

\ifthenelse{\boolean{uprightparticles}}%
{

 \def\Pmu         {\ensuremath{\upmu}\xspace}

 \def\Ppsi        {\ensuremath{\uppsi}\xspace}

 \def\PDelta      {\ensuremath{\Delta}\xspace}                 
 \def\PXi      {\ensuremath{\Xi}\xspace}                 
 \def\PLambda      {\ensuremath{\Lambda}\xspace}                 
 \def\PSigma      {\ensuremath{\Sigma}\xspace}                 
 \def\POmega      {\ensuremath{\Omega}\xspace}                 
 \def\PUpsilon      {\ensuremath{\Upsilon}\xspace}                 
 
 \def\PB      {\ensuremath{\mathrm{B}}\xspace}                 
                  
 \def\PD      {\ensuremath{\mathrm{D}}\xspace}

 \def\PH      {\ensuremath{\mathrm{H}}\xspace}                 
                  
 \def\PJ      {\ensuremath{\mathrm{J}}\xspace}                 
 \def\PK      {\ensuremath{\mathrm{K}}\xspace}

 \def\PW      {\ensuremath{\mathrm{W}}\xspace}

 \def\PZ      {\ensuremath{\mathrm{Z}}\xspace}                 
                  
 \def\Pb      {\ensuremath{\mathrm{b}}\xspace}                 
 \def\Pc      {\ensuremath{\mathrm{c}}\xspace}                 
                  
 \def\Pe      {\ensuremath{\mathrm{e}}\xspace}

 \def\Pi      {\ensuremath{\mathrm{i}}\xspace}

 \def\Pt      {\ensuremath{\mathrm{t}}\xspace}

}
{

 \def\Pmu         {\ensuremath{\mu}\xspace}

 \def\Ppsi        {\ensuremath{\psi}\xspace}                 
                  
 \mathchardef\PDelta="7101
 \mathchardef\PXi="7104
 \mathchardef\PLambda="7103
 \mathchardef\PSigma="7106
 \mathchardef\POmega="710A
 \mathchardef\PUpsilon="7107
                  
 \def\PB      {\ensuremath{B}\xspace}                 
                  
 \def\PD      {\ensuremath{D}\xspace}

 \def\PH      {\ensuremath{H}\xspace}                 
                  
 \def\PJ      {\ensuremath{J}\xspace}                 
 \def\PK      {\ensuremath{K}\xspace}

 \def\PW      {\ensuremath{W}\xspace}

 \def\PZ      {\ensuremath{Z}\xspace}                 
                  
 \def\Pb      {\ensuremath{b}\xspace}                 
 \def\Pc      {\ensuremath{c}\xspace}                 
                  
 \def\Pe      {\ensuremath{e}\xspace}

 \def\Pi      {\ensuremath{i}\xspace}

 \def\Pt      {\ensuremath{t}\xspace}

}

\makeatletter
\ifcase \@ptsize \relax
  \newcommand{\miniscule}{\@setfontsize\miniscule{4}{5}}
\or
  \newcommand{\miniscule}{\@setfontsize\miniscule{5}{6}}
\or
  \newcommand{\miniscule}{\@setfontsize\miniscule{5}{6}}
\fi
\makeatother

\DeclareRobustCommand{\optbar}[1]{\shortstack{{\miniscule (\rule[.5ex]{1.25em}{.18mm})}
  \\ [-.7ex] $#1$}}

\def\en         {{\ensuremath{\Pe^-}}\xspace}   
\def\ep         {{\ensuremath{\Pe^+}}\xspace}

\def\mup        {{\ensuremath{\Pmu^+}}\xspace}
\def\mun        {{\ensuremath{\Pmu^-}}\xspace} 

\def\H      {{\ensuremath{\PH^0}}\xspace}

\def\W      {{\ensuremath{\PW}}\xspace}
\def\Wp     {{\ensuremath{\PW^+}}\xspace}
\def\Wm     {{\ensuremath{\PW^-}}\xspace}

\def\cquark    {{\ensuremath{\Pc}}\xspace}
\def\cquarkbar {{\ensuremath{\overline \cquark}}\xspace}
\def\ccbar     {{\ensuremath{\cquark\cquarkbar}}\xspace}
\def\bquark    {{\ensuremath{\Pb}}\xspace}
\def\bquarkbar {{\ensuremath{\overline \bquark}}\xspace}
\def\bbbar     {{\ensuremath{\bquark\bquarkbar}}\xspace}
\def\tquark    {{\ensuremath{\Pt}}\xspace}
\def\tquarkbar {{\ensuremath{\overline \tquark}}\xspace}
\def\ttbar     {{\ensuremath{\tquark\tquarkbar}}\xspace}

\def\wpb {{\ensuremath{\W\!+\!\bbbar}}\xspace}
\def\wppb {{\ensuremath{\Wp\!+\!\bbbar}}\xspace}
\def\wmpb {{\ensuremath{\Wm\!+\!\bbbar}}\xspace}

\def\wpc {{\ensuremath{\W\!+\!\ccbar}}\xspace}
\def\wppc {{\ensuremath{\Wp\!+\!\ccbar}}\xspace}
\def\wmpc {{\ensuremath{\Wm\!+\!\ccbar}}\xspace}

\def\hbb {\decay{\H}{\bbbar}}
\def\hcc {\decay{\H}{\ccbar}}
\def\antikt     {\ensuremath{\text{anti-}k_{\mathrm{T}}}\xspace}

\def\ugb       {\mbox{\textsc{uGB}}\xspace}

  \def\Kbar    {{\kern 0.2em\overline{\kern -0.2em \PK}{}}\xspace}

\def\KorKbar    {\kern 0.18em\optbar{\kern -0.18em K}{}\xspace}

  \def\Dbar    {{\kern 0.2em\overline{\kern -0.2em \PD}{}}\xspace}

\def\DorDbar    {\kern 0.18em\optbar{\kern -0.18em D}{}\xspace}

\def\Bbar    {{\ensuremath{\kern 0.18em\overline{\kern -0.18em \PB}{}}}\xspace}

\def\BorBbar    {\kern 0.18em\optbar{\kern -0.18em B}{}\xspace}

\def\jpsi     {{\ensuremath{{\PJ\mskip -3mu/\mskip -2mu\Ppsi\mskip 2mu}}}\xspace}

  \def\Y#1S{\ensuremath{\PUpsilon{(#1S)}}\xspace}

\def\Lbar        {{\ensuremath{\kern 0.1em\overline{\kern -0.1em\PLambda}}}\xspace}
\def\LorLbar    {\kern 0.18em\optbar{\kern -0.18em \PLambda}{}\xspace}

\def\BF         {{\ensuremath{\cal B}}\xspace}

\def\BR         {\BF}
\newcommand{\decay}[2]{\ensuremath{#1\!\to #2}\xspace}         

\def\to                 {\ensuremath{\rightarrow}\xspace}

\def\AT#1     {\ensuremath{A_{\mathrm{T}}^{#1}}\xspace}           

\def\C#1      {\ensuremath{\mathcal{C}_{#1}}\xspace}                       
\def\Cp#1     {\ensuremath{\mathcal{C}_{#1}^{'}}\xspace}                    
\def\Ceff#1   {\ensuremath{\mathcal{C}_{#1}^{\mathrm{(eff)}}}\xspace}        
\def\Cpeff#1  {\ensuremath{\mathcal{C}_{#1}^{'\mathrm{(eff)}}}\xspace}       
\def\Ope#1    {\ensuremath{\mathcal{O}_{#1}}\xspace}                       
\def\Opep#1   {\ensuremath{\mathcal{O}_{#1}^{'}}\xspace}                    

\newcommand{\tev}{\ifthenelse{\boolean{inbibliography}}{\ensuremath{~T\kern -0.05em eV}\xspace}{\ensuremath{\mathrm{\,Te\kern -0.1em V}}}\xspace}
\newcommand{\gev}{\ensuremath{\mathrm{\,Ge\kern -0.1em V}}\xspace}
\newcommand{\mev}{\ensuremath{\mathrm{\,Me\kern -0.1em V}}\xspace}
\newcommand{\kev}{\ensuremath{\mathrm{\,ke\kern -0.1em V}}\xspace}
\newcommand{\ev}{\ensuremath{\mathrm{\,e\kern -0.1em V}}\xspace}
\newcommand{\gevc}{\ensuremath{{\mathrm{\,Ge\kern -0.1em V\!/}c}}\xspace}
\newcommand{\mevc}{\ensuremath{{\mathrm{\,Me\kern -0.1em V\!/}c}}\xspace}
\newcommand{\gevcc}{\ensuremath{{\mathrm{\,Ge\kern -0.1em V\!/}c^2}}\xspace}
\newcommand{\gevgevcccc}{\ensuremath{{\mathrm{\,Ge\kern -0.1em V^2\!/}c^4}}\xspace}
\newcommand{\mevcc}{\ensuremath{{\mathrm{\,Me\kern -0.1em V\!/}c^2}}\xspace}

\def\invfb   {\ensuremath{\mbox{\,fb}^{-1}}\xspace}

\def\gsim{{~\raise.15em\hbox{$>$}\kern-.85em
          \lower.35em\hbox{$\sim$}~}\xspace}
\def\lsim{{~\raise.15em\hbox{$<$}\kern-.85em
          \lower.35em\hbox{$\sim$}~}\xspace}

\def\pt         {\mbox{$p_{\rm T}$}\xspace}

\def\pythia     {\mbox{\textsc{Pythia}}\xspace}

\def\tell1  {TELL1\xspace}
\def\ukl1   {UKL1\xspace}

